\documentstyle[aps,preprint,epsf]{revtex} 
\def\btt{\begin{tabular}{ll}} 
\def\ett{\end{tabular}} 
\def\et{\end{tabbing}} 
\def\ieee#1#2#3{{IEEE Trans. Inf. Theor.} {\bf #1}, #2 (#3)} 
\def\bpj#1#2#3{{Biophys. J.} {\bf #1}, #2 (#3)} 
 
\def\complex#1#2#3{{Complexity} {\bf #1}, #2 (#3)} 
 
\def\jcb#1#2#3{{J. Comp. Biol.} {\bf #1}, #2 (#3)} 
\def\statsc#1#2#3{{Stat. Sci.} {\bf #1}, #2 (#3)} 
 
\def\prl#1#2#3{{ Phys. Rev. Lett.} {\bf #1}, #2 (#3)} 
\def\pre#1#2#3{Phys. Rev. E {\bf #1}, #2 (#3)} 
\def\pra#1#2#3{Phys. Rev. A {\bf #1}, #2 (#3)}

\def\ijbc#1#2#3{Int. J. Bifurcation and Chaos {\bf #1}, #2 (#3)} 
\def\physd#1#2#3{Physica D {\bf #1}, #2 (#3)} 
\def\physa#1#2#3{Physica A {\bf #1}, #2 (#3)} 
\def\nat#1#2#3{Nature {\bf #1}, #2 (#3)} 
\def\epl#1#2#3{Europhys. Letts.{\bf #1}, #2 (#3)}

\def\u{\underline}
 
\def\etl{$et ~al.$~} 
\def\uu{{\it U. urealyticum~}}

\def\beq{\begin{equation}} 
\def\bc{\begin{center}} 
\def\ec{\end{center}} 
\def\eqn{\end{equation}\noindent} 
\def\eqnr{\end{eqnarray}} 
\def\beqr{\begin{eqnarray}\noindent}

\def\S{{\cal S}} 
\def\F{{\cal F}} 
\begin{document} 
\title{Simplifying the mosaic description of DNA sequences} 
\author{Rajeev K. Azad\thanks{Present address: School of Biology, 
Georgia Institute of Technology, Atlanta, GA 30332, USA}, 
J. Subba Rao, 
Wentian Li$^1$, and 
Ramakrishna Ramaswamy 
\thanks{For correspondence, email:{\tt rama@vsnl.com}} $^2$ 
} \address{School of Environmental 
Sciences, $^2$School of Physical Sciences\\ Jawaharlal Nehru 
University, New Delhi 110 067, India\\ $^1$ 
Center for Genomics and Human Genetics, North Shore - LIJ Research 
Institute,\\ 
Manhasset, NY 11030, USA\\} 

\date{\today} 
\maketitle 
\begin{abstract} 
By using the Jensen-Shannon divergence, genomic DNA can be divided into 
compositionally distinct domains through a standard recursive 
segmentation 
procedure. Each domain, while significantly different from its 
neighbours, may however share compositional similarity with one or more 
distant (non--neighbouring) domains. We thus obtain a coarse--grained 
description of the given DNA string in terms of a smaller set of 
distinct domain labels. This yields a {\it minimal} domain description 
of a given DNA sequence, significantly reducing its organizational 
complexity. This procedure gives a new means of evaluating 
genomic complexity as one examines organisms ranging from bacteria to 
human. The mosaic organization of DNA sequences could have originated 
from 
the {\it insertion} of fragments of one genome (the parasite) inside 
another (the host), and we present numerical experiments that are 
suggestive of this scenario. 
\end{abstract} 
\noindent 
\noindent 

\section{INTRODUCTION} 

One of the major goals in DNA sequence analysis is in gaining an
understanding of the overall organization of the genome.  Beyond
identifying the manifestly functional regions such as genes, promoters,
repeats, etc., it has also been of interest to analyse the properties 
of
the DNA string itself.  One set of studies has been directed towards
examining the nature of correlations between the bases.  There is some
evidence for long-range correlations which give rise to {\it 1/f} 
spectra
in genomic DNA \cite{li4,li1,li5}; this feature has been attributed to 
the
presence of {\it complex heterogeneities} in nucleotide sequences
\cite{li5}.  These result in hierarchical patterns in DNA, the mosaic 
or
`domain within domain' picture \cite{pedro2}.  This structure is most
conveniently explored through segmentation analysis based on 
information
theoretic measures \cite{pedro2,pedro3,pedro1,li6}, although other 
schemes
to uncover the correlation structure over long scales, such as 
detrended
fluctuation analysis of DNA walks \cite{peng} or wavelet tranform 
technique
\cite{audit} have also been applied.  There have been some attempts to
decode the biological implications of such complexity
\cite{audit,li2,li3}, but these are incompletely understood as of now.  
On
shorter length scales there is a prominent 3-base correlation in coding
regions of DNA; this offers a means of locating and identifying genes
\cite{shrish}.  There are other short--range correlations as well
\cite{widom,trifonov} corresponding to structural constraints on the 
DNA
double helix.

Segmentation analysis is a powerful means of examining the large--scale
organization of DNA sequences
\cite{pedro2,pedro3,pedro1,abrs,ramen,braun,pedro4}.  The most commonly
used procedure \cite{pedro2,pedro3,pedro1} is based on maximization of 
the
Jensen-Shannon (J-S) {\it divergence} through which a given DNA string 
is
recursively separated into compostionallly homogeneous segments called
domains (or patches).  This results in a coarse-grained description of 
the
DNA string as a sequence of distinct domains.  The criterion for 
continuing
the segmentation process is based on statistical
significance (this is equivalent to hypothesis testing)
\cite{pedro2,pedro3} or, alternatively, within a model selection 
framework
based on the Bayesian information criterion \cite{li6}.  This
criterion can be extended and used to detect isochores \cite{li6}, CpG
islands, origin and terminus of replication in bacterial genomes, 
complex
repeats in telomere sequences, etc.\cite{li7}.  Segmentation using a
12-symbol alphabet derived from codon usage has been shown recently to
delineate the border between coding and noncoding regions in a DNA 
sequence
\cite{pedro1}.

In the present work, we analyse the segmentation structure of genomic 
DNA
for a class of genomes ranging in (evolutionary) complexity from 
bacteria
to human.  Our motivation is to understand the complexity of genome
organization in terms of the domains obtained.  We further aim to 
correlate
the domain picture with evolutonary biological processes.

By construction a given domain is heterogenous with respect to its
neighbours, but it may nevertheless be compositionally similar to other
domains.  Based on this premise, we attempt to draw a larger domain 
picture
by obtaining `domain sets'.  These consist of a set of domains which 
are
homogeneous when concatenated.  A domain set may thus be interpreted as 
a
larger homogeneous sequence, parts of which are scattered nonuniformly 
in a
genomic sequence.  The number of domain sets constructed thus is found 
to
be much fewer than the domains obtained upon segmentation
\cite{pedro2,pedro3,pedro1,li6}.  We propose here an optimal procedure,
starting from the domains found from one of the above segmentation 
methods,
and building up a domain set by adding together all its components.  We
then use standard complexity measures to show that this gives a 
superior
model in as much as the complexity is reduced.

This paper is organised as follows.  In the next section, we briefly 
review
the segmentation methods based on the J-S divergence.  Section III 
contains
our main results.  We first segment a given genome to reveal the 
primary
domain structure that derives from the J-S divergence.  We then show 
how
the domain sets are constructed, and analyse the attendant decrease in
complexity.  In Section IV, we speculate that such domain organization
ocurred during genomic evolution when there was lateral gene and/or DNA
transfer between species.  To that end, we present the results of 
numerical
experiments based on a host-parasite model, where we artificially 
insert
fragments of one genome inside another, and demonstrate that this 
process
can be uncovered via segmentation.  Section V concludes the paper with 
a
summary and discussion of our results.

\section{Segmentation methods} 
In this section we briefly review the segmentation methodology that is 
used 
here in order to fragment a genome into homogeneous domains. Consider 
a sequence $\S$ as a concatenation of two subsequences 
$\S^{(1)}$ and $\S^{(2)}$. The 
Jensen--Shannon divergence \cite{lin} of the subsequences is 
\beqr 
\label{jsd} 
{\bf D}(\F^{(1)}, \F^{(2)}) &=& H(\pi^{(1)} \F^{(1)}+\pi^{(2)} 
\F^{(2)}) 
-[\pi^{(1)} H(\F^{(1)})+\pi^{(2)} H(\F^{(2)})], 
\eqnr 
where $\F^{(i)}=\{f_1^{(i)},f_2^{(i)},...,f_k^{(i)}\}, i=1,2$ are the 
relative frequency vectors, and 
$\pi^{(1)}$ and $\pi^{(2)}$ their weights. In Eq.~(\ref{jsd}), 
$H$ is the Shannon 
entropy (in unit of bits) 
\beq 
H(\F)=- \sum_{i=1}^k f_i~ \log_2~ f_i, 
\eqn 
although, as can be appreciated, a variety of other functions on the 
$f_i$'s can also be used as a criterion for estimating the divergence 
of 
two sequences. 

The algorithm proposed by Bernaola-Galv\'{a}n \etl \cite{pedro2,pedro3} 
proceeds as follows. 
A sequence is segmented in two domains such that the J-S divergence 
$\bf D$ 
is maximum over all possible partitions. Each resulting domain is then 
further 
segmented recursively. 

The main issue with regard to continual segmentation is that unless the 
significance of a given segmentation step is properly assessed, it is 
possible to arrive at segments which have no great significance. 
This question is also related to a second issue, namely when one should 
stop 
the recursion. Since we consider finite DNA sequences, it is again 
possible to keep segmenting until the segments are very short. Both 
these 
questions can be answered through one of two possible approaches which 
we 
now describe. 

\subsection{Hypothesis testing framework} 

The statistical significance of the segmentation is determined by 
computing 
the maximum value of the J-S divergence for the two potential 
subsegments, ${\bf D}_{max}$, and estimating the probability of getting 
this value or less in a random sequence. This defines 
the significance level, $s(x)$, as 
\beq 
s(x)=Prob\{{\bf D}_{max} \leq x \}. 
\eqn 
The probablility distribution of ${\bf D}_{max}$ has an analytic 
approximation \cite{pedro3,pedro1} and 
\beq 
s(x) = [F_{\nu} (\beta \cdot 2N\ln 2 \cdot x)]^{N_{eff}}, 
\eqn 
where $F_{\nu}$ is the chi--square distribution function with $\nu$ 
degrees 
of freedom, $N$ is the sequence length, $\beta$ is a scale factor which 
is essentially 
independent of $N$ and $k$ and for each $k$, $N_{eff} = a~\ln N +b.$ 
The 
values of $\beta$ and $N_{eff}$ (and thus the constants $a$ and $b$) 
are 
found from Monte Carlo simulations by fitting the empirical 
distributions 
to the above expression \cite{pedro3,pedro1}. 

Within the hypothesis testing framework, then, the segmentation is 
allowed if and only if $s(x)$ is greater than a preset level of 
statistical significance. It is possible to segment a given sequence 
initially at a (usually very high) significance level, and these 
domains 
are further segmented at lower levels of significance to detect the 
inner 
structure or other patterns \cite{abrs}.

\subsection{Model selection framework} 

A different criterion can be evolved for 
stopping the recursive segmentation within the so-called model 
selection 
framework \cite{li6}. This is based on the Bayesian information 
criterion 
\cite{jeffreys,rafter,hastie}, denoted ${\bf B}$ below,
\beq 
\label{bic} 
{\bf B }= -2 \log(\hat{L})+ \log(N)K + O(1)+ O(\frac{1}{\sqrt{N}})+ 
O(\frac{1}{N}), 
\eqn 
where $\hat{L}$ is the maximum likelihood of the model, $N$ is the 
sample 
size and $K$ is the number of parameters in the model. 

A potential segmentation based on the J-S divergence $\bf D$ is deemed 
acceptable if ${\bf B}$ is reduced after segmentation. From the above 
equation, this condition is \cite{li6} 

\beq 
2N {\bf D}>(K_{2}-K_{1}) \log(N), 
\label{wenti} 
\eqn 
where $K_1$ and $K_2$ are the number of free parameters of the models 
before and after the segmentation. This is the lower bound of the 
significance level; an upper bound can be preset by using 
a measure of {\it segmentation strength} \cite{li6}, 
\beq 
s={{2N {\bf D}-(K_{2}-K_{1}) \log(N)} \over {(K_{2}-K_{1}) \log(N)}}. 
\eqn 
Eq.~(\ref{wenti}) is equivalent to the condition $s>0$. 

\section{Applications and analysis} 

In the present work we consider DNA sequences as strings in a 4--letter 
alphabet ($A,T,C, G$). In the model selection framework 
discussed above, therefore, the relevant parameters are $K_1=3$ (since 
only 
3 of the 4 nucleotides are independent) and $K_2=7$ (the 3 free 
parameters 
from each of the two subsegments, and in addition, the partition point 
which is another independent parameter) \cite{li6}. The importance of 
this segmentation approach in detecting some of the structural and 
functional units in DNA sequences has been demonstrated recently 
\cite{li7}. 
The results that follow have been obtained by the application of this 
approach.

\subsection{Labeling the domains} 

The complete genome of a bacterium {\it Ureaplasma urealyticum} (751719 
bp)
and a contig of human chromosome 22 ({\it gi $\mid 10879979 \mid$ ref 
$\mid
NT\_011521.1 \mid$}, 767357 bp) were segmented at the lower bound of 
the
stopping criterion, namely Eq.~(\ref{wenti}).  The number of segments
obtained by this procedure is 86 for the bacterium and 248 for human
chromosome 22 contig.  Labeling each of these segments by a unique 
symbol
gives a coarse--grained view of the entire sequence, say $S_1\cdot
S_2\cdots S_N$.

While each segment $S_k$ is heterogeneous with respect to its 
neighbours,
$S_{k\pm 1}$, it need not be compositionally distinct from a
non--neighbouring segment, $S_j$.  Therefore, we now examine the {\it 
inter
se} heterogeneity of all segments with respect to each other.  Segments
$S_k$ and $S_j$ are concatenated, and if this `supersegment' cannot be
segmented by the same criterion, then both $S_k$ and $S_j$ are assigned 
the
same domain symbol.  This is done recursively and exhaustively, so that
within the model selection framework of segmentation, all domains that
cannot be distinguished from one another are assigned the same symbol. 
This gives a reduced and further coarse--grained view of the domain
structure of a DNA sequence.

To ensure that the above procedure is as complete and self--consistent 
as
possible, we examine each segment $S_k$ by concatenating it with $S_j$ 
and
\underline{all} preceding distinct segments that share the same domain
symbol as $S_j$, and examine whether this larger sequence can be 
segmented. 
Explicitly, if segments $S_i$ and $S_j$ have the same symbol (following 
the
procedure given above) we examine the supersegment $S_i\cdot S_j\cdot 
S_k$
to determine whether segment $S_k$ should share the same domain symbol 
or
not.  It is further required to to consider all possible subsets 
($S_i\cdot S_k$, $S_j\cdot S_k$, etc.) to ensure that all segments that 
are 
deemed to share a given domain symbol do indeed belong to one class, 
namely 
that such superdomains do not undergo further segmentation.

Following the above, the 86 domains obtained from the 
segmentation of \uu are reduced to a total of 17 distinct domain types: 

$\begin{array}{cccccccccccc} 
\u{S_1} & S_2 & S_3 & S_4 & S_5 & S_3 & \u{S_1} & S_2 & \u{S_1} & S_6 & 
S_4 & 
\u{S_1}\\ 
S_6 & S_7 & S_2 & \u{S_1} & S_6 & S_4 & S_8 & S_9 & S_4 & S_9 & S_{10} 
& S_4\\ 
S_9 & S_4 & S_{11} & S_{12} & S_6 & S_4 & S_{10} & S_6 & S_{10} & S_6 & 
S_{11} & S_6\\ 
S_7 & S_6 & S_{11} & S_7 & S_3 & S_{11} & S_3 & S_{10} & S_6 & S_3 & 
S_9 & S_{11}\\ 
S_{10} & S_4 & S_{11} & S_{10} & S_{13} & S_4 & S_{13} & S_9 & S_{11} & 
S_4 & S_6 & S_4\\ 
S_{11} & S_4 & S_{14} & S_6 & S_8 & S_6 & S_{14} & S_4 & S_6 & S_{15} & 
\u{S_1} & S_9\\ 
S_4 & S_{16} & S_9 & S_{17} & S_{15} & S_6 & S_{17} & S_7 & S_{17} & 
\u{S_1} & S_{17} & S_8\\ 
S_{16} & S_{14}\\ 
\end{array}$ 

The 248 segments of human chromosome 22 also undergo simplification, to 
a 
total of 53 distinct domain types: 

$\begin{array}{cccccccccccc} 

S_1 & S_2 & S_3 & S_4 & S_5 & S_4 & S_3 & S_6 & S_4 & S_6 & S_7 & S_4\\ 
S_8 & S_4 & S_9 & S_{10} & S_6 & S_4 & S_7 & S_1 & S_4 & S_7 & S_6 & 
S_4\\ 
S_7 & S_{11} & S_4 & S_{12} & \u{S_{13}} & S_4 & S_{14} & S_{12} & S_4 
& S_{15} & S_{16} & S_{14}\\ 
S_6 & S_9 & S_{10} & S_{17} & S_{16} & S_{10} & S_{16} & S_6 & S_{12} & 
S_{18} & S_{12} &S_{10}\\ 
S_3 & S_1 & S_3 & S_1 & S_{10} & S_9 & S_6 & S_3 & S_{12} & S_{16} & 
S_3 & S_{12}\\ 
S_{14} & S_1 & S_7 & S_6 & S_{12} & S_7 & S_1 & S_6 & S_{19} & S_6 & 
S_{20} & S_{17}\\ 
S_7 & S_{21} & S_7 & S_{22} & S_{21} & S_{22} & S_{23} & S_7 & S_{23} & 
S_{24} & S_{17} & S_{21}\\ 
S_7 & S_{21} & S_1 & S_{21} & S_7 & S_{21} & S_7 & S_{16} & S_{25} & 
S_1 & S_{16} & S_{15}\\ 
S_{26} & S_8 & S_{15} & S_8 & S_{21} & S_8 & S_{21} & S_{27} & S_{16} & 
S_{12} & S_1 & S_{28}\\ 
S_{21} & S_{28} & S_{21} & S_{12} & S_{21} & S_{16} & S_{12} & S_{16} & 
S_{12} & S_{28} &S_{16}&S_{19}\\ 
S_{17} & S_{27} & S_{28} & S_{16} & S_{20} & S_{21} & S_{29} & S_{25} & 
S_{30} & S_{25}&S_{31}& S_{25}\\ 
S_{28} & S_8 & S_{25} & S_{29} & S_{32} & S_3 & S_{25} & S_{31} & 
S_{33} & S_8 & S_{31} & S_{34}\\ 
S_{31} & S_{29} & S_{30} & S_{31} & S_{35} & S_{36} & S_{21} & S_{36} & 
S_{37} & S_{36} & S_2&S_{36}\\ 
S_9 & S_1 & S_9 & \u{S_{13}} & S_{38} & \u{S_{13}} & S_{39} & S_{29} & 
S_{34} & S_{37} & S_2 & S_{29}\\ 
S_{40} & S_{41} & S_{31} & S_{37} & S_{31} & \u{S_{13}} & S_{35} & 
S_{42} & S_9 & S_5 & S_9 & S_{42}\\ 
S_7 & S_{41} & S_1 & S_{43} & S_{44} & S_{45} & S_{46} & S_{42} & 
S_{45} & S_{47} & S_{45} &S_{44}\\ 
S_{32} & S_{44} & S_{45} & S_{44} & S_{48} & S_{43} & S_{25} & S_{45} & 
S_{11} & 
S_{49} &\u{S_{13}}&S_{49}\\ 
S_{11} & S_{49} & S_{47} & S_{50} & S_{47} & \u{S_{13}} & S_{26} & 
\u{S_{13}} & 
S_{44} & \u{S_{13}} &S_{45}&\u{S_{13}}\\ 
S_8 & S_9 & S_{45} & S_{50} & S_9 & S_{51} & S_5 & S_{52} & S_{32} & 
S_{51} & S_5 & S_{51}\\ 
S_{45} & S_9 & S_{21} & S_2 & S_9 & S_{21} & S_9 & S_{39} & S_9 & 
S_{43} & 
\u{S_{13}} & S_{53}\\ 
S_{39} & \u{S_{13}} & S_{43} & \u{S_{13}} & S_{49} & \u{S_{13}} & 
S_{47} & \u{S_{13}} \\
\end{array}$ 

This gives a maximally coarse--grained view of the DNA squence, in 
terms of 
``domain sets'': these are the elements of a given domain type which 
may be 
scattered over the entire genome. Examples above are domains like $S_1$ 
in 
bacterium or $S_{13}$ in human which are widely dispersed (these are 
underlined for visual clarity above), suggesting that 
these fragments possibly had a common origin, or that they were 
inserted at 
the same time during evolution. Expansion--modification 
\cite{lipra,ohno} and 
insertion--deletion \cite{buld} are thought to play major role in 
evolution: 
the former ensures duplication accompanied by point mutations in 
genomes 
and the latter results in insertion of a part of chromosome inside a 
nucleotide sequence or deletion of base pairs from a nucleotide 
sequence. 
An initial homogeneous sequence may thus become heterogeneous by 
insertions/deletions that consistently go on with the evolution. 
Insertions may cause the pieces of a homogeneous sequence to spread.

\subsection{Insertion--deletion and heterogeneity} 

The process of insertion--deletion \cite{buld} has played an important 
role in increasing the complexity of genomes. Motivated by the 
simplification 
of domain description as above, 
we perform the following numerical experiment in order to examine 
the increase in complexity by such processes. Fragments of the 
\uu bacterial sequence of total length $80$ Kbp are inserted at $N$ 
random 
positions in the human chromosome 22 contig 
($gi|10879979|ref|NT\_011521.1|$). The heterogeneity 
will naturally increase because of such insertions. 

Prior to the insertion of bacterial fragments, the total number of 
domains 
in the human chromosome 22 contig is 248; after inserting the fragments 
at random 
positions, in a typical realization, the number of segments obtained is 
375. The results of such experiments can be quantified through the 
sequence compositional complexity \cite{pedro4,pedro5}, denoted 
${\bf S}$,
\beqr 
\label{scc} 
{\bf S}&=&H(S)-\sum_{i=1}^{n} {\frac {n_{i}}{N}} H(S_i) \nonumber\\ 
&=&\sum_{i=1}^{n} {\frac{n_{i}}{N}}[H(S)-H(S_i)], 
\eqnr 
where $S$ denotes the whole sequence of length $N$ and $S_i$ is the 
$i$th 
domain of length $n_i$. This measure, which is independent of the 
length 
of sequence quantifies the difference or dispersion among the
compositions of the domains. The higher the ${\bf S}$, the more 
heterogeneous the DNA sequence. 

When fragments of very different composition are inserted into a given 
DNA 
sequence, the complexity will necessarily increase. We compute 
$\Delta_S = 
{\bf S'-\bf S}$ for domains obtained after and before the insertion for 
the  example as above and also for a number 
of genomes. In all cases $\Delta_S>0$: the compositional complexity 
increases after insertion. If deletion is also introduced, say by 
removing a fragment of random length from a random position (the range 
of 
lengths being deleted is kept same as that of the `inserts') in general 
$\Delta_S$ increases further. 

\subsection{Measuring the complexity} 
We quantify the simplication of domain description of the two 
representative 
genomes by considering a complexity measure within the model selection 
framework, 
namely the Bayesian information criterion (${\bf B}$). Within standard 
statistical analyisis, one model is superior in comparison with another 
if 
it has a lower ${\bf B}$. For the case of \uu, where the segmentation 
procedure gives 86 domains, 
\beqr 
{\bf B}&=& -2 \log(\hat{L})+ 343\log(N) 
\eqnr 
where $K=343$ parameters correspond to $86 \times 3$ base compositions 
and $85$ borders. 
These are reorganized 
into 17 domain sets, and thus \beqr 
{\bf B'}&=& -2 \log(\hat{L^{\prime}})+136\log(N) 
\eqnr 
($136=17\times 3+85$). 
The maximum likelihood can be expressed as 
\beq 
L(p_{\alpha})=\prod_{\alpha} p_{\alpha}^{N_{\alpha}}, 
\eqn 
where $\{p_{\alpha}\}$ and $\{N_{\alpha}\}$ are the base composition 
parameter and the base counts respectively corresponding to alphabet 
$\{\alpha=A,T,G,C\}$ of a sequence. $\Delta_B={\bf B'}-{\bf B}$ 
depends on the relative contribution of both terms; typically 
$L> L'$ since the first segmentation uses a more accurate 
measurement of base composition. The reduction in this measure comes from 
the second term through the drastic reduction in the number of domains 
which reduces the model complexity. 

For {\it U. urealyticum} and human, $\Delta_B = -1709$ and $-4884$ 
respectively which shows that the model representative of the 
domain set is better 
than the original one (we use the lower bound i.e. $\Delta_B<0$ for 
determining the statistical significance \cite{li6}). As another 
example, we 
found $\Delta_B$ for {\it Thermoplasma acidophilum} (archaeabacteria, 
$1564906$ bp) and another contig of human chromosome 22 ({\it gi $\mid 
10880022 \mid$ ref $\mid NT\_011522.1 \mid$}, $1528072$ bp) to be 
$-2808$ 
and $-10420$ respectively. We repeated this procedure for different 
available genomes and found the above results to be consistent. Note 
that 
the simplication can also be quantified in terms of ${\bf S}$ and we 
observe 
$\Delta_S<0$ in all cases. 

\section{{\it In silico} experiments on domain insertion: a 
host--parasite 
perspective} 

It is tempting to speculate that the heterogeneity that is uncovered by 
the 
segmentation procedures discussed above is a reflection of the 
evolutionary 
history of the given sequence, and in particular, that the different 
domains arise from insertion processes acting at different evolutionary 
times. For instance, it is well--known that the human genome contains a 
small fraction of bacterial genome which have most likely arisen from 
processes such as viral insertion or lateral gene transfer. 

To what extent can the segmentation process determine the exact pattern 
of 
insertions? Here we describe some simple experiments that are designed 
to 
explore this question. Starting with a homogeneous fragment of human 
DNA, we insert fragments from (a homogeneous segment of) bacterial 
genomes; 
this increases the heterogeneity. We then apply the segmentation 
algorithm 
followed by the labeling procedure and compare the results with the 
(known) control. 

Experiments were done on a homogeneous domain set from the human 
genome, of 
total length 100139 bp. Into this, fragments from a homogeneous segment 
of 
length 17584 bp from the genome of \uu were inserted. In a 
representative 
case, we took 3 fragments (of lengths 5000, 7000 and 5584 bp 
respectively) 
and inserted them at locations 10000, 50000 and 92000 in the human 
genome 
domain. 

Upon segmentation, all seven segments were identified, with the 
boundaries 
between the bacterial and human DNA sequences determined as follows: 
9984 (10000), 15000 (15000), 49751, 50060 (50000), 
56968 (57000), 91636 (92000) and 97575 (97584), (the exact values are 
given in brackets). There is thus one false 
positive, but otherwise all the boundaries are determined to fairly 
high 
precision. The domain sets can also be reconstructed, and the seven 
segments, $\S_1\S_2\S_1\S_2\S_1\S_2\S_1$ conform to two sets. 

Shown in Fig.~1(a) is the insertion process for a case where fragments 
from 
two bacterial genomes, {\it Ureaplasma urealyticum} and {\it 
Thermoplasma 
acidophilum} are randomly inserted in the human genome segment. 
Carrying 
out segmentation at varying strength $s$ gives a greater number of 
segments 
compared to the correct value of 13. With $s=0.2$, one gets 18 segments 
(see Fig.~1(b)) which is the best reconstruction possible within the 
present framework. On obtaining domain sets, we find that up to about 
85\% 
of human and \uu genomes are properly identified, the errors affecting 
the reconstruction of {\it T. acidophilum} which is only 67\% accurate. 

To summarize, our results from several numerical experiments show that 
the 
reconstruction of the fragmentation process can be done to high 
accuracy so 
long as the inserted fragments are sufficiently long and widely 
separated. 

\section{Discussion and Summary} 

Segmentation offers a novel view of the compositional heterogeneity of 
a
DNA sequence.  In the present work we have applied the segmentation
analysis to genomic sequences from several organisms.

Our main focus has been on understanding the organization and to this 
end
we have applied a number of different analytical tools.  Our main 
analysis
has been directed towards obtaining a coarse--grained representation of 
DNA
as a string of minimal domain labels.  Complexity measures indicate 
that
the reduced model in terms of domain sets is superior to a model where 
each
domain is treated as independent.

Insofar as the different domains are considered, our main hypothesis is
that these arise when fragments of one (possibly homogenous) DNA 
sequence
get randomly inserted into another (also possibly homogenous) sequence.  
A
controlled set of (numerical) experiments give support to this 
hypothesis:
we are able to identify domain boundaries to high accuracy so long as
inserted domains are not very short.  The accuracy could be further
increased by improving the segmentation process, for example, using 1 
to 3
segmentation rather than the binary or 1 to 2 segmentation used here:
binary segmentation is only one of several possible segmentation 
procedures (see
Ref.~\cite{braun}).
 
A consequence of this analysis, and one that we are currently 
exploring, is 
that different domains (or domain sets) in one genome can have 
arisen via insertion from another organism. Homology analysis (say by 
the 
use of standard tools such as BLAST or FASTA) can help to unravel the 
origins of the domains. Thus segmentation analysis can possibly help in 
reconstructing the evolutionary history of the genome. 

\section*{ACKNOWLEDGMENT:} 
RR is supported by a grant from the Department of BioTechnology, India. 

\newpage 
\begin{figure}
\vskip1cm
\epsfxsize=3.0in
\epsfbox[20 120 350 450]{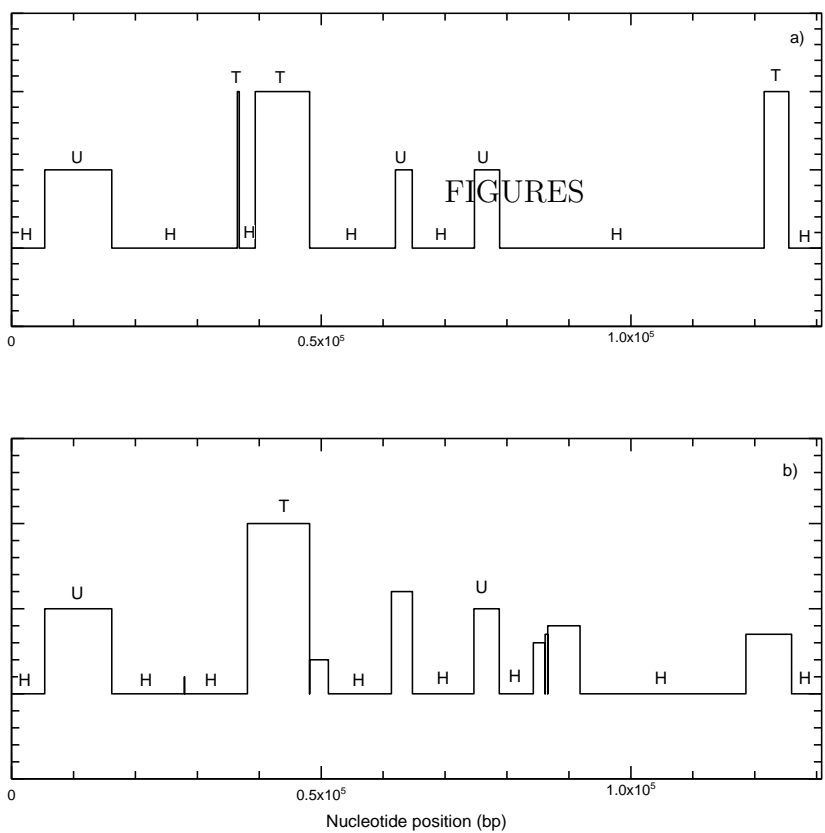}
\vskip1cm
\caption{(a) Representation of a DNA sequence obtained by random 
insertion of fragments of two bacterial sequences {\it T. acidophilum} 
(T) and 
{\it U. urealyticum} (U) into a human sequence (H) (see text). 
(b) The domain structure as uncovered by the procedure of 
segmentation and labeling (as described in the text).} 
\end{figure}

\end{document}